\documentclass[a4paper,11pt]{article}
\usepackage{jinstpub} % for details on the use of the package, please see the JINST-author-manual
\usepackage{booktabs}
\usepackage{lineno}
\usepackage{makecell}
\usepackage{siunitx}
\DeclareSIUnit\year{yr}

\title{Effect of increased \textit{DCR} on the detection of minimum-ionizing particles with SiPMs}

\author[1]{K. Neumann \note{Corresponding author.}}
\author{, M. Antonello,}
\author{L. Brinkmann,}
\author{E. Garutti}
\author{and J. Schwandt}

\affiliation{University of Hamburg, Institute for Experimental Physics\\Luruper Chaussee 149, 22761 Hamburg, Germany}

\emailAdd{katjana.neumann@uni-hamburg.de}

\abstract{Radiation damage to a Silicon Photomultiplier (SiPM), as it occurs during the lifetime of the planned CMS high-granularity calorimeter detector, increases the dark current and degrades the signal-to-noise separation for minimum-ionizing particles (MIPs) and their detection efficiency. To investigate these effects, a plastic scintillator tile air-coupled to a SiPM is used to detect MIPs from a  \(^{90}\mathrm{Sr}\) source, in a single-channel design similar to the tiles of the CMS high-granularity calorimeter upgrade.
We compared the SiPM responses after actual radiation exposure with responses simulated in the laboratory by increasing the dark-count rate (\textit{DCR}) through optical illumination with an LED light source. This optical method induces no structural damage or deep defects, thus isolating the effect of increased dark-count rate. Our results show that both radiation-induced damage and LED-induced dark-count rate increases lead to similar reductions in the MIP signal and the signal-to-noise ratio. This indicates that the primary factor for the performance degradation is the elevated dark-count rate itself, rather than additional defects in the silicon. The results demonstrate that the key effects of radiation damage on SiPMs can be effectively replicated using controlled optical illumination, providing a practical and easily accessible approach for evaluating and optimizing SiPM performance under radiation-like conditions in the laboratory.
}


\begin{document}
\maketitle
\flushbottom

\section{Introduction}
\label{sec:intro}
Silicon Photomultipliers (SiPMs) are pixel arrays of avalanche photodiodes that operate above the breakdown voltage and are widely used as photon detectors. They offer numerous advantages, including the ability to detect single photons, robustness, compactness, high photon detection efficiency, and insensitivity to magnetic fields. However, SiPMs also have drawbacks such as a high dark-count rate (\textit{DCR}) at room temperature, correlated noise from crosstalk and after-pulses, and radiation sensitivity. Since SiPMs are often deployed in harsh radiation environments, such as particle accelerators, it is essential to study the effects of radiation damage on their performance. A relevant example is the CMS high-granularity calorimeter (HGCAL), a high-precision sampling calorimeter composed of silicon and scintillator modules \cite{CERN-LHCC-2017-023}. At the High-Luminosity LHC SiPMs in the HGCAL will be exposed to radiation conditions with fluences up to \(\Phi_\mathrm{eq}\sim\SI{5e13}{\centi\meter^{-2}}\) \cite{HGCAL_scint_sipm}. Parts of this calorimeter use the SiPM-on-tile technology pioneered by the CALICE collaboration \cite{Sefkow_2019}, with more than 200,000 SiPMs \cite{HGCAL}. Here, the minimum-ionizing particle (MIP) response is used as a standard candle for the calibration of the single channels, which requires a signal-to-noise (\(S/N\)) separation above 2.5 during the whole lifetime of the detector at an operation overvoltage of \(\Delta V=1\)-\(\SI{1.5}{\volt}\) and at a temperature of \(T=\SI{-35}{\celsius}\). Radiation damage affects the \(S/N\), which decreases with irradiation fluence due to several effects. Most prominent is the SiPM's \textit{DCR} increase \cite{CENTISVIGNALI2018137}, due to increasing leakage or dark current, which also increases the SiPMs self-heating \cite{selfheating}. Another effect is the gain degradation with increasing fluence. For example, a reduction in gain of \(20\%\) can be observed for a fluence of \(\Phi_\mathrm{eq}=\SI{5e13}{\centi\meter^{-2}}\) \cite{radhardsinglecell}. The reduction in gain and also in photon-detection efficiency (\(PDE\)) leads to a significant decrease of the signal, which can reach \(\approx40\%\) for high fluences as \(\Phi_\mathrm{eq}=\SI{2e14}{\centi\meter^{-2}}\) \cite{Musienko_2020}. Further studies show that irradiation leads to an increase of the breakdown voltage, \(V_\mathrm{bd}\) \cite{Musienko_2020}, and of the turn-off voltage, \(V_\mathrm{off}\), by the same amount \cite{Raduniforminsipm}. In this context, the breakdown voltage refers to the voltage at which avalanche multiplication sets in, and the turn-off voltage refers to the voltage at which avalanche multiplication stops. To operate the SiPM at a fixed voltage above \(V_\mathrm{bd}\) can compensate this effect. This paper presents a method to compare the SiPM's MIP response under increased \textit{DCR} caused either by true radiation damage or by DC triggered LED illumination. While radiation damage induces structural changes, the LED method increases the dark current exclusively through illumination, without inducing defects. The goal is to demonstrate that mimicking radiation-induced \textit{DCR} increase with LED illumination can produce comparable effects, providing a simple way to reproduce radiation damage phenomena in laboratory conditions.
In Section~\ref{sec:setup} the development of a method to compare the SiPM response to MIPs under increased \textit{DCR} caused either by radiation damage or LED illumination is presented. Section~\ref{sec:results} analyzes the correlation between MIP-response, \textit{DCR} and \(S/N\) ratio, including experimental results and their implications. Finally, the paper concludes with a discussion on optimizing operational conditions to mitigate performance degradation under high radiation fluences.

\section{Setup and Method}
\label{sec:setup}
\begin{figure}[htbp]
    \centering
    \includegraphics[width=0.5\linewidth]{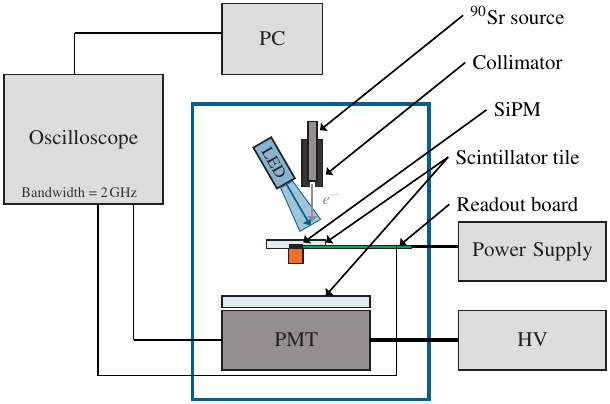}
    \vspace*{2mm}
    \caption{Schematic illustration of the setup used to measure the SiPMs response to MIPs \cite{MTKatjana}.}
    \label{fig:light_yield_setup}
\end{figure}

To measure the MIP response and noise of a tile-SiPM system the setup shown in Figure~\ref{fig:light_yield_setup} is used. The whole setup is contained in a light tight box and consists out of an Hamamatsu MPPC S14160-9766 SiPM (Table~\ref{tab:Technical_specifications_of_the_used_SiPMs}), which is coupled to a \(30\times30\times3\,\SI{}{\metre\metre^3}\) Bicron 408 scintillator tile. 

A readout board is positioned inside the box and directly connected to the SiPM, the SiPM's power supply (Keithley 2470 Source Meter) and the RTO64 Rohde \(\&\) Schwarz \(\SI{2}{\giga\hertz}\) oscilloscope, with a resolution of \(\SI{100}{\pico\second}\), that is used for the data acquisition. 
The customized readout board AC couples the SiPM signal with a capacitance of \(\SI{1}{\nano\farad}\) and has a noise filter to prevent current spikes from the power supply. In addition, the readout board includes an effective \(\SI{2}{\kilo\ohm}\) resistor, which leads to a voltage drop between power supply and SiPM. This voltage drop is small (<\(\mathcal{O}(\mu\mathrm{V})\)) for the non-irradiated SiPM, but become large (\(\mathcal{O}(V)\)) for the SiPM irradiated at the highest fluence. The SiPM is also connected via a copper finger to a copper plate, cooled using a LAUDA ECO RE1050 chiller. Below the SiPM and the cooling system a scintillator tile read out via a PhotoMultiplier Tube (PMT, Hamamatsu R5900U-00-M4) is placed, which is used as a trigger and powered by a high voltage source (Ortec Model 401B). The PMT is connected to the oscilloscope. For the MIP response measurements a \(^{90}\)Sr source with an activity of \(\SI{2}{\mega Bq}\) is positioned above the SiPM. A DC triggered blue LED (NSPB500AS, \(\SI{470}{\nano\meter}\)) is used to mimic the \textit{DCR} increase induced by irradiation of the SiPM. For this a DC current is provided by a Yokogawa Gs200 current source to the LED.

\begin{figure}[htbp]
    \centering
    \includegraphics[width=0.99\textwidth]{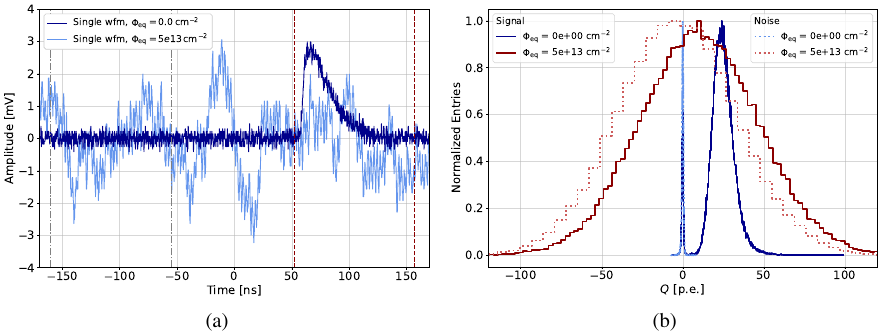}
    \caption{a) The single baseline-subtracted waveforms for the non-irradiated (dark blue) and highest irradiated (\(\Phi_\mathrm{eq}=\SI{5e13}{\centi\meter^{-2}}\)) SiPM (light blue) for \(V_\mathrm{ex,set}=\SI{3}{\volt}\).  
    The grey and dark red dashed lines indicate the pedestal and signal region, respectively. b) The corresponding integrals.}
    \label{fig:non_highest_fluence_3V_wfm_hists}
\end{figure}

\begin{table}[htbp]
    \centering
    \begin{tabular}{cccccccc}
        \toprule
        \(N_\mathrm{pix}\)&pixel size [\(\mu\)m]&\(A_\mathrm{eff}\) [\(\mathrm{mm}^2\)]& \(\lambda_\mathrm{p}\) [nm]&\(V_\mathrm{bd}\) [V]&\(V_\mathrm{off}\) [V]&\(\tau\)[ns]& \(\Phi_\mathrm{eq}\) [\(\mathrm{cm}^{-2}\)]\\
        \midrule
        8480&\(15\) & \(1.4\times 1.4\) & 480 & \(36.68\)& \(35.92\) & \(17.51\)& \(0.0\), \(2\cdot 10^{12}\),  \\
        &&&&\(\pm\,0.01\)&\(\pm\,0.01\)&\(\pm\,0.03\)&\(1\cdot 10^{13}\), \(5\cdot 10^{13}\) \\
        \bottomrule
    \end{tabular}
    \caption{Technical specifications of the used SiPM Hamamatsu MPPC S14160-9766 at \(T=\SI{-20}{\celsius}\). \\\(N_\mathrm{pix}\): number of pixels, \(A_\mathrm{eff}\): effective photo sensitive area, \(\lambda_\mathrm{p}\): peak sensitivity wavelength, \(V_\mathrm{bd}\): breakdown voltage, \(V_\mathrm{off}\): turn-off voltage and \(\tau\): decay time.}
    \label{tab:Technical_specifications_of_the_used_SiPMs}
\end{table}In this study, a new, non-irradiated SiPM was used, along with three neutron-irradiated SiPMs with \(\SI{1}{\mega\electronvolt}\) neutron equivalent fluences ranging from \(\Phi_\mathrm{eq}=\SI{2e12}{\centi\meter^{-2}}\) to \(\SI{5e13}{\centi\meter^{-2}}\) (see Table~\ref{tab:Technical_specifications_of_the_used_SiPMs}).
The SiPMs are operated at \(T=\SI{-20}{\celsius}\) and at a set excess voltage of \(V_\mathrm{ex,set}=V_\mathrm{bias}-V_\mathrm{bd}=2-\SI{4}{V}\) above the breakdown voltage \(V_\mathrm{bd}\). The current versus time of each event is recorded as waveform. Each waveform is integrated for two separate intervals of equal length. The first ends \(\approx\SI{110}{\nano\second}\) before the trigger to light pulse and is used to measure the baseline and noise of the system. The second gate (signal region) starts \(\sim \SI{3}{\nano\second}\) before the pulse (see Figure~\ref{fig:non_highest_fluence_3V_wfm_hists}a). Both gates are \(t_\mathrm{gate}=\SI{105}{\nano\second}\) long, corresponding to the integration of the full signal. The resulting charge integrals \(Q\) [\(\SI{}{\volt\second}\)] are stored as uncorrelated binned arrays. 
\(Q\) is then converted into the number of photo electrons, by dividing by the charge of a single discharge for the non-irradiated SiPM in the absence of recovery effects:
\begin{equation}
    Q[\mathrm{p.e.}]=\frac{Q[\mathrm{V\,s}]}{q_0[\mathrm{C}]\cdot G\cdot R[\Omega]}
\end{equation}
where \(q_0\) is the elementary charge, \(G\) the gain of the SiPM and \(R=\SI{50}{\ohm}\) the oscilloscope’s load resistance. 
From the charge distribution in the pedestal region (see Figure~\ref{fig:non_highest_fluence_3V_wfm_hists}b) the noise, i.e.\>pedestal width, is extracted using a gaussian fit. The MIP, corresponding to the most probable value (mpv) of the histogram of the charge in the signal region, is extracted from these histograms using a fit of a Landau distribution convolved with a Gaussian provided by the 'pylandau' Python package \cite{pylandau}. 
\section{Results}
\label{sec:results}
\begin{figure}[htpb]
    \centering
    \includegraphics[width=0.98\linewidth]{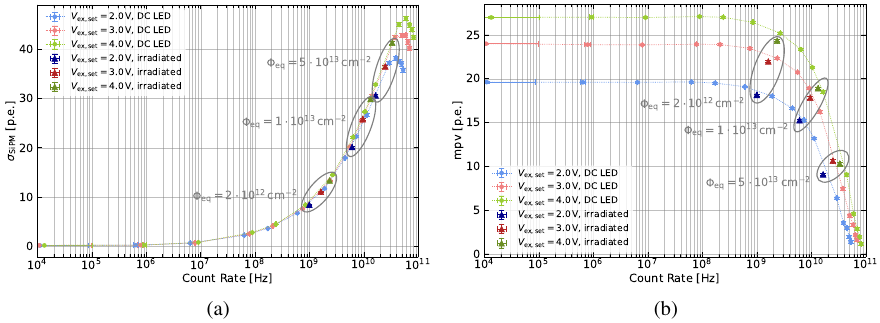}
    \caption{a) The SiPMs noise (\(\sigma_\mathrm{SiPM}\)) and b) the MIP response as a function of the dark-count rate for excess voltages, \(V_\mathrm{ex,set}\), between \(\SI{2}{\volt}\) and \(\SI{4}{\volt}\) at \(T=\SI{-20}{\celsius}\).}
    \label{fig:sig_sipm_vs_dcr}
\end{figure}

The SiPMs noise, \(\sigma_\mathrm{SiPM}\) is shown in Figure~\ref{fig:sig_sipm_vs_dcr} as a function of the dark-count rate, including correlated noise \(CN\), determined from the dark current \(I_\mathrm{dark}\) by
\begin{equation}
    DCR^{*}=DCR(1+CN)=\frac{I_\mathrm{dark}}{q_0\cdot G}
\end{equation}
for different DC LED illumination, irradiation fluences and three different excess voltages. 

\begin{figure}[htbp]
    \centering
    \includegraphics[width=\linewidth]{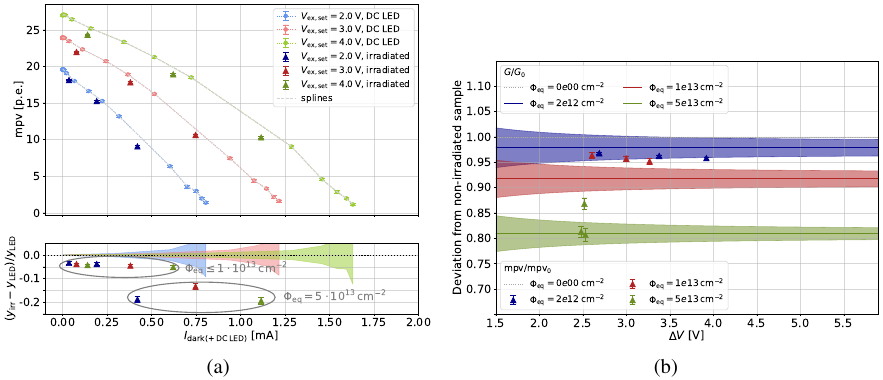}
    \caption{a) The most-probable value (MIP response) as a function of the (dark-)current \(I_\mathrm{dark(+\,DC\:LED)}\) for excess voltages, \(V_\mathrm{ex,set}\), of \(\SI{2}{\volt}\) to \(\SI{4}{\volt}\) at \(T=\SI{-20}{\celsius}\). Splines are fitted to the data (gray) and deviations of irradiated samples from these splines are shown (lower part Figure~a). b) Comparison of the reduction in gain \cite{radhardsinglecell} and deviation of the MIP response from values of the non-irradiated sample for different fluences. Adapted from \cite{MTKatjana}.}
    \label{fig:mpv_vs_idark}
\end{figure}

\begin{figure}[h!]
    \centering
    \includegraphics[width=\linewidth]{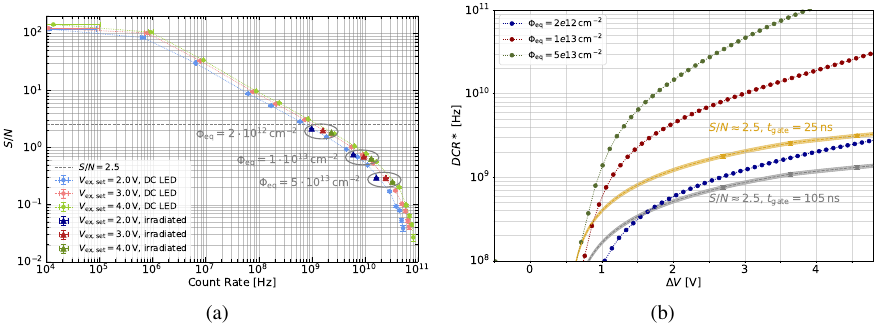}
    \caption{a) The signal to noise ratio (\(S/N\)) as a function of the dark-count rate (\textit{DCR}) for excess voltages of \(V_\mathrm{ex,set}=2\)-\(\SI{4}{\volt}\) at \(T=\SI{-20}{\celsius}\). The grey dashed line corresponds to \(S/N=2.5\). b) \textit{DCR} vs. overvoltage \(\Delta V\). The dashed line indicates the \textit{DCR} at which \(S/N=2.5\), for \(t_\mathrm{gate}=\SI{105}{\nano\second}\) (grey) and \(\SI{25}{\nano\second}\) (yellow). Points below the dashed lines have a higher \(S/N\) than 2.5 in the respective cases.}
    \label{fig:sn_ratio_vs_dcr_voltage}
\end{figure}

The results indicate that the increase of \(\sigma_\mathrm{SiPM}\) with increasing \textit{DCR} is visible for both the irradiated and the non-irradiated SiPMs, where the \textit{DCR} is emulated by the random photons of the DC illumination. Figure~\ref{fig:mpv_vs_idark}b shows the MIP response as a function of the dark-count rate. The MIP response increases with voltage as expected due to higher gain and \textit{PDE}. Furthermore, it decreases with increasing dark current or fluence. 
To compare the data obtained from DC illumination used to emulate irradiation and the truly irradiated samples, a spline is fitted through the former data (see Figure~\ref{fig:mpv_vs_idark}a). This spline serves as a reference to quantify the differences in response between irradiated SiPMs and those subjected to emulated irradiation (lower part Figure~\ref{fig:mpv_vs_idark}a).
A deviation of \(\lesssim 5\%\) for the fluences of \(\Phi_\mathrm{eq}=\SI{2e12}{\centi\meter^{-2}}\) and \(\SI{1e13}{\centi\meter^{-2}}\) and a deviation of \(\sim 20\%\) for the highest fluence of \(\Phi_\mathrm{eq}=\SI{5e13}{\centi\meter^{-2}}\) can be observed.
From the work, described in \cite{radhardsinglecell}, in which a SiPM of the same pixel size and series was used, it is known that the gain reduces with increasing fluence. A comparison of the deviation of the mpv of irradiated samples from the non-irradiated sample with DC illumination replicating the same dark current as for the irradiated ones and the reduction in gain for different fluences (data taken from Table~3, in \cite{radhardsinglecell}) is shown in Figure~\ref{fig:mpv_vs_idark}b. The ratio of the irradiated and non-irradiated samples is plotted as a function of the overvoltage \(\Delta V=V_\mathrm{bias}-V_\mathrm{off}\). A comparable reduction in gain and mpv is observed for the lowest and highest fluence investigated. 
For the fluence of \(\Phi_\mathrm{eq}=\SI{1e13}{\centi\meter^{-2}}\), a reduction in the MIP response of approximately \(4\)-\(5\%\) is observed, whereas the reduction in gain, according to \cite{radhardsinglecell}, is \(8\)-\(9\%\). 
 This leads to the conclusion that the difference between the data taken with irradiated samples and the data taken with DC illumination can be exclusively explained by the reduction in gain due to the irradiation.\\
\newline
Figure~\ref{fig:sn_ratio_vs_dcr_voltage}a shows the \(S/N\) ratio as a function of the dark-count rate. As expected a decrease of the \(S/N\) ratio with increasing count rate is visible and a slight voltage dependence is observed. Further, for count rates larger than \(\approx\SI{750}{\mega\hertz}\), which here corresponds to irradiation fluences larger than \(\Phi_\mathrm{eq}=\SI{2e12}{\centi\meter^{-2}}\) and excess voltages of \(V_\mathrm{ex,set}=2-\SI{4}{\volt}\) the signal-to-noise ratio is smaller than 2.5. 

As mentioned in Section 1, the CMS HGCAL requires this separation at overvoltages of \(\Delta V=1-\SI{1.5}{\volt}\) for channel calibration throughout operation. Therefore, strategies to enhance the \(S/N\) ratio for the irradiated samples have to be considered.
An increase of the \(S/N\) ratio can for instance be achieved by reducing the voltage, the integration gate and/or the temperature. Since the temperature was fixed at \(T=\SI{-20}{\celsius}\) in this study, only the first two options can be examined. Figure~\ref{fig:sn_ratio_vs_dcr_voltage}b displays the \(DCR\) as a function of the applied overvoltage. 
In addition, the voltage and \(DCR\) dependent \(S/N\) limit value of \(2.5\) is shown in for integration gates of \(t_\mathrm{gate}=\SI{105}{\nano\second}\) (grey) and \(t_\mathrm{gate}=\SI{25}{\nano\second}\) (yellow). For this \(S/N\) value and \(t_\mathrm{gate}=\SI{105}{\nano\second}\), the \(DCR\) values for the respective voltages (\(V_\mathrm{ex,set}=2,\,3\>\mathrm{and}\>\SI{4}{\volt}\)) are taken from Figure~\ref{fig:sn_ratio_vs_dcr_voltage}a with an uncertainty of \(5\%\) (shaded band). The behavior of the \(S/N\) value over the voltage range shown here is extrapolated from these points using a spline (dashed lines). 

From Figure~\ref{fig:sn_ratio_vs_dcr_voltage}b it is visible that reducing \(t_\mathrm{gate}\) rises the \textit{DCR} threshold at which \(S/N<2.5\). In addition, it can be observed that for the lowest fluence of \(\SI{2e12}{\centi\meter^{-2}}\) and \(\Delta V=1-\SI{1.5}{\volt}\) the \(S/N\) ratio is larger than \(2.5\). 
For the two highest fluences, the voltage is at the lowest limit of \(\Delta V=\SI{1}{\volt}\). Decreasing the overvoltage further, however, introduces a strong voltage dependence of the response. To achieve the required \(S/N\) ratio across all investigated fluences, the temperature must be reduced further, as it is indeed the case for the CMS HGCAL, where the proposed operating temperature is \(T=\SI{-35}{\celsius}\) \cite{HGCAL}.

\section{Conclusion}
\label{sec:conclusion}
The MIP response and \(S/N\) ratio for a SiPM-on-tile system are characterized before and after irradiation with reactor neutrons up to a \(\SI{1}{\mega\electronvolt}\) neutron equivalent fluence of \(\SI{5e13}{\centi\meter^{-2}}\). For a fluence of \(\SI{1e13}{\centi\meter^{-2}}\) the \(S/N\) ratio for MIP detection is greater than 2.5 if operating the SiPM slightly below \(\Delta V=\SI{1}{\volt}\) at a temperature of \(\SI{-20}{\celsius}\) and an integration length of \(t_\mathrm{gate}=\SI{25}{\nano\second}\). Further cooling would enable the selection of an overvoltage above \(\SI{1}{\volt}\), where the response voltage dependence is smaller.
The presented method of emulating radiation damage through DC LED illumination provides a straightforward approach to investigate the impact of increased dark-count rate on the SiPM performance. 
Our comparison between irradiated SiPMs and non-irradiated samples demonstrates, that both methods yield consistent results when accounting for known radiation damage induced gain reduction. The study confirms that the primary effect of radiation damage on the MIP response and noise characteristics can be effectively reproduced using controlled illumination, enabling more accessible laboratory testing and optimization studies on operating conditions. With this method, we identify operational limits influenced by signal-to-noise ratio considerations, highlighting the importance of optimizing temperature, voltage and integration length conditions to maintain performance at higher fluences. Overall, this approach provides a practical tool to evaluate the behavior of SiPMs under radiation-like conditions, reducing the need for detailed irradiation studies.

\acknowledgments
We acknowledge the support from BMBF via the project 05H24GUB (High-D-Calo). This
work is supported by the Deutsche Forschungsgemeinschaft (DFG, German Research Foundation) under Germany’s Excellence Strategy, EXC 2121, Quantum Universe (390833306). \vspace{0.4cm}

\noindent\textbf{\large{Data Avaliability Statement}} \vspace{0.4cm}\newline
This article has associated data in a data repository. The data and
software associated with this study are openly available at: \cite{dataset} and \cite{software}.

\end{document}